\documentclass[aps,prd,floatfix,12pt]{revtex4}
\usepackage{epsfig}
\usepackage{bm}
\usepackage{dcolumn}
\usepackage{latexsym}

\begin{document}

\title{Thermodynamics of lattice QCD with 3 flavours of colour-sextet quarks II:
$\bm{N_t=6}$ and $\bm{N_t=8}$.}

\author{J.~B.~Kogut}
\affiliation{Department of Energy, Division of High Energy Physics, Washington,
DC 20585, USA}
 \author{\vspace{-0.2in}{\it and}}
\affiliation{Dept. of Physics -- TQHN, Univ. of Maryland, 82 Regents Dr.,
College Park, MD 20742, USA}
\author{D.~K.~Sinclair}
\affiliation{HEP Division, Argonne National Laboratory, 9700 South Cass Avenue,
Argonne, IL 60439, USA}

\begin{abstract}

We have been studying QCD with 2 flavours of colour-sextet quarks as a
candidate walking-Technicolor theory using lattice-QCD simulations. The
evolution of the coupling constant with lattice spacing is measured at the
finite-temperature chiral transition to determine if this theory is
asymptotically free and hence QCD-like. The lattice spacing is varied by
changing the number of lattice sites, $N_t$, in the Euclidean time direction.
QCD with 3 flavours is studied for comparison. Since this theory is expected
to be conformal, with an infrared fixed point, the coupling constant at the
chiral transition should approach a non-zero value as $N_t$ becomes large. Our
earlier simulations on lattices with $N_t=4$ and $N_t=6$ exhibited a
significant decrease in coupling at the chiral transition as $N_t$ was
increased. We have now extended these simulations to $N_t=8$, and performed
additional simulations at $N_t=6$ to measure the coupling constant at the
chiral transition more precisely. These indicate that while there is an
appreciable decrease in coupling between $N_t=6$ and $N_t=8$, this is much
smaller than that between $N_t=4$ and $N_t=6$. Thus we are hopeful that we are
approaching the large-$N_t$ limit. However, further simulations at larger
$N_t$(s) are needed.

\end{abstract}

\maketitle

\section{Introduction}

The LHC has demonstrated its ability to probe the Higgs sector of the Standard
Model. This has revived interest in trying to understand this sector in more
detail. It is understood that the original description of this part of the 
standard model in terms of an elementary scalar field is at best an effective 
field theory. We are interested in studying models where the Higgs field is
composite. The most promising of these are the Technicolor theories
\cite{Weinberg:1979bn,Susskind:1978ms}, Yang-Mills
gauge theories with massless fermions whose Goldstone (techni-)pions play the
role of the Higgs field giving masses to the $W^\pm$ and $Z$. It has been
observed that phenomenological difficulties with Technicolor theories which are
simply scaled-up QCD can be avoided if the fermion content is chosen such that
there is a range of mass scales over which the gauge coupling evolves very
slowly -- `walks'
\cite{Holdom:1981rm,Yamawaki:1985zg,Akiba:1985rr,Appelquist:1986an}. 

For a chosen gauge group and $N_f$ fermions in a chosen representation, there
is typically a value of $N_f$ below which the one- and two-loop terms in 
$\beta$, which describes the evolution of the coupling are both negative, and
the theory is asymptotically free and probably also confining with spontaneously
broken chiral symmetry (QCD-like). For $N_f$ large enough both terms are
positive, and asymptotic freedom is lost. In between these regimes is a range of
$N_f$s for which the 1-loop term is negative and the 2-loop term is positive.
Theories in this regime are still asymptotically free, but if this two-loop
$\beta$ is a good guide to its behaviour, there is a second, infrared-%
attractive fixed point, and the theory is conformal. However, if before this
would-be IR fixed point is reached, the coupling becomes strong enough that
a chiral condensate forms, this fixed point is avoided and the theory is 
confining. Because of the nearby IR fixed point, there will be a range of
mass scales over which the coupling evolves very slowly, and the theory walks.
For an extensive discussion of this behaviour for $SU(N)$ gauge theories and
a guide to the earlier literature, see for example \cite{Dietrich:2006cm}.
Such behaviour is clearly non-perturbative, and lends itself to study using
simulation methods developed for lattice QCD.

We \cite{Kogut:2010cz,Kogut:2011ty,Sinclair:2013era}, 
along with others \cite{Shamir:2008pb,DeGrand:2008kx,DeGrand:2009hu,%
DeGrand:2010na,DeGrand:2012yq,DeGrand:2013uha,Fodor:2008hm,Fodor:2011tw,%
Fodor:2012uu,Fodor:2012uw,Fodor:2012ty,Fodor:2014pqa},
have been studying (techni)-QCD with 2 flavours of
(techni-)colour-sextet (techni)-quarks, which is a candidate 
walking-Technicolor theory. For the rest of the paper we will drop the prefix
'techni' and remember that we are dealing not with normal QCD, but with a
scaled-up version where $f_\pi \approx 246$~GeV, rather than $\approx 93$~MeV.
For QCD with colour-sextet quarks, the 3-flavour theory is also in the regime
where the 1- and 2-loop contributions have opposite sign. 3 flavours is so
close to the number ($3\frac{3}{10}$) above which asymptotic freedom is lost,
that the non-trivial zero of the 2-loop $\beta$-function is at a small enough
value of $\alpha_s=g^2/4\pi$ ($\approx 0.085$), that it is unlikely that a
chiral condensate is formed before it is reached. Thus it is believed that
this theory will have an infrared fixed point at a value close to that
predicted perturbatively, and so for massless quarks, it will be conformal.
We therefore find it useful to study QCD with 3 colour-sextet quarks for 
comparison with QCD with 2 colour-sextet quarks, to see if they do indeed 
behave rather differently. In fact non-lattice methods, which attempt to
determine the walking and conformal windows, place the critical number of 
flavours for QCD with colour-sextet quarks closer to 2 
\cite{Dietrich:2006cm,Appelquist:1988yc,Sannino:2004qp,Poppitz:2009uq,%
Armoni:2009jn,Ryttov:2007cx,Antipin:2009wr}. 

We simulate lattice QCD with 3 flavours of light colour-sextet staggered quarks
at finite temperatures using the RHMC method \cite{Clark:2006wp} to tune to 3
flavours. The simple Wilson plaquette action is used for the gauge fields.
Finite temperature is achieved by simulating on a lattice whose spatial extent
$N_s$ is infinite and whose temporal extent $N_t$ is finite. In practice,
$N_s$ is also finite, but $N_s >> N_t$. The temperature is then $T=1/N_ta$
where $a$ is the lattice spacing. If $T$ is fixed at a finite temperature
phase transition, then $a \rightarrow 0$ as $N_t \rightarrow \infty$. We work
with the chiral-symmetry restoration transition (which means that we need to
extrapolate to zero quark mass). If this is indeed a finite temperature
transition, the coupling constant at this transition will approach zero as
$N_t \rightarrow \infty$ in a manner described by asymptotic freedom. In the
case where the continuum theory is conformal and chiral symmetry is unbroken
in the continuum limit, chiral symmetry is restored at a bulk transition,
where the coupling constant is finite and independent of $N_t$ if $N_t$ is
sufficiently large.

Our earlier simulations of the 3-flavour theory were performed at $N_t=4$ and
$6$ \cite{Kogut:2011bd}. Between $N_t=4$ and $6$, the value of $\beta=6/g^2$ 
at the chiral transition ($\beta_\chi$) showed a relatively large increase 
($\sim 0.3$). This we interpreted as being because it occurs at a coupling 
which is strong enough that the fermions are bound into a chiral condensate and
do not contribute significantly to the evolution of the coupling constant. The
theory is effectively a pure gauge theory, and what is observed is the
finite-temperature chiral transition of the quenched theory where
coupling-constant evolution is that of 0-flavour QCD. We have now extended our
simulations to $N_t=8$. (Preliminary results were presented at Lattice 2012
and Lattice 2013 \cite{Sinclair:2013era}.) In addition we have had to extend
our $N_t=6$ simulations to use more $\beta$ values close to the chiral
transition. Between $N_t=6$ and $8$ $\beta_\chi$ increases by $\sim 0.1$ which
is small enough to indicate that its evolution is no longer controlled by
quenched dynamics, however, it does not yet indicate that the coupling is
approaching a finite constant. Thus larger $N_t$s are called for.

In section~2 we give some technical details. Section~3 describes the simulations
and results. Discussions and conclusions are given in section~4.

\section{Technical details}

We use the unimproved Wilson (plaquette) action for the gauge fields:
\begin{equation}
S_g=\beta \sum_\Box \left[1-\frac{1}{3}{\rm Re}({\rm Tr}UUUU)\right].
\end{equation}
The unimproved staggered-fermion action is used for the sextet quarks:
\begin{equation}
S_f=\sum_{sites}\left[\sum_{f=1}^{N_f/4}\psi_f^\dagger[D\!\!\!\!/+m]\psi_f
\right],
\end{equation}
where $D\!\!\!\!/ = \sum_\mu \eta_\mu D_\mu$ with 
\begin{equation}
D_\mu \psi(x) = \frac{1}{2}[U^{(6)}_\mu(x)\psi(x+\hat{\mu})-
                            U^{(6)\dagger}_\mu(x-\hat{\mu})\psi(x-\hat{\mu})].
\end{equation}
Here $U^{(6)}$ is the sextet representation of $U$, i.e. the symmetric part of
the tensor product $U \otimes U$. When $N_f$ is not a multiple of $4$ we 
use the fermion action:
\begin{equation}
S_f=\sum_{sites}\chi^\dagger\{[D\!\!\!\!/+m][-D\!\!\!\!/+m]\}^{N_f/8}\chi.
\end{equation}
The operator which is raised to a fractional power is positive definite and
we choose the real positive root. This yields a well-defined operator. We 
assume that this defines a sensible field theory in the zero lattice-spacing
limit, ignoring the rooting controversy. (See for example \cite{Sharpe:2006re}
for a review and guide to the literature on rooting.) What is important for
comparison with the 2-flavour theory is that this action differs only in the
value of $N_f$ from the action used in those simulations.

We simulate using the Rational Hybrid Monte Carlo (RHMC) algorithm, in which 
the fractional powers of the quadratic Dirac operator are approximated by an
optimal (diagonal) rational approximation, to the required precision. A global
Metropolis accept/reject step at the end of each trajectory removes errors due
to discretization of molecular-dynamics time, associated with updating the 
fields.

If the massless theory is QCD-like -- confining with spontaneously broken chiral
symmetry -- in the continuum limit, the chiral phase transition and the 
deconfinement transition (to the extent that it is well-defined) will be finite
temperature transitions occurring at fixed temperatures $T_\chi$ and $T_d$
respectively. At small enough lattice spacing -- large enough $N_t$ -- the
evolution of the lattice coupling constants will be governed by the 
Callan-Symanzik $\beta$-function. To 2-loops:
\begin{equation}
\beta(g) = -b_1 g^3 - b_2 g^5.
\end{equation}
Expressing our coupling constant evolution in terms of $\beta=6/g^2$ (We
apologize for the fact that we are using $\beta$ for 2 different purposes)
\begin{equation}
\Delta\beta(\beta) = \beta(a) - \beta(\lambda a)
                   = (12b_1 + 72b_2/\beta)\ln(\lambda)
\label{eqn:deltabeta}
\end{equation}
to this order in $g$. For $N_f$ flavours of sextet quarks,
\begin{eqnarray}
b_1 &=& \left(11 - \frac{10}{3}N_f\right)/16\pi^2 \nonumber \\
b_2 &=& \left(102 - \frac{250}{3}N_f\right)/(16\pi^2)^2 \; .
\end{eqnarray}
To this order the second zero of $\beta$, the one associated with an IR fixed
point occurs at $g=g_c$, where:
\begin{equation}
g_c^2 = -b_1/b_2
\end{equation}
For $N_f=3$, this gives
\begin{equation}                                                              
g_c^2 = 4\pi^2/37 \approx 1.067 \; ,
\end{equation} 
and thus
\begin{equation}
\alpha_s = g_c^2/(4\pi) \approx 0.0849 \; .
\end{equation}
At this coupling the lattice quantity
\begin{equation}
\beta=\beta_c=6/g_c^2 \approx 5.623 \; .
\end{equation}
Note that to this order, the $\beta$-function has the same coefficients 
independent of renormalization scheme, and whether we are referring to a 
`bare' or to a `renormalized' coupling. So, this zero also occurs at the same
value, independent of scheme.

Since these two-loop calculations indicate that the running coupling does not
get very large before the infrared fixed point is reached, it would be 
surprising if a condensate formed to enable this fixed point to be avoided. We
contrast this with the 2-flavour case, where $g_c^2$ and hence $\alpha_s$ is an
order of magnitude larger than for the 3-flavour theory. Approximate
calculations and arguments support the idea that QCD with 3 sextet quarks is
conformal \cite{Dietrich:2006cm,Appelquist:1988yc,Sannino:2004qp,%
Poppitz:2009uq,Armoni:2009jn,Ryttov:2007cx,Antipin:2009wr}

The values of $\beta_d$ at $N_t=4$ and $6$ are much smaller than those of
$\beta_\chi$ at the same $N_t$s. Since we have concluded that the chiral
transition for these $N_t$ values lies in the strong-coupling regime where 
coupling constant evolution is described by quenched dynamics, the deconfinement
transition lies at even stronger couplings and is unlikely to emerge until much
larger $N_t$. In fact, $\beta_d$ lies considerably below $\beta_c$ at these
$N_t$s, a region inaccessible to 2-loop perturbation theory. We therefore 
concentrate our efforts on the evolution of the coupling constant at the chiral
transition.

Although the chiral condensate, measured in our simulations, shows evidence that
it will vanish in the chiral limit for sufficiently large $\beta$, its $\beta$
dependence for the quark masses we use is too smooth to enable an accurate
determination of $\beta_\chi$. This is still true when we use the method of
the Lattice Higgs Collaboration \cite{Fodor:2011tw} to remove much of the 
$a^{-2}$ divergence from the lattice quantity. We therefore use the position of
the peak in the chiral susceptibility extrapolated to $m=0$ as our estimate of
$\beta_c$. The chiral susceptibility is given by
\begin{equation}
\chi_{\bar{\psi}\psi} = V\left[\langle(\bar{\psi}\psi)^2\rangle
                      -        \langle\bar{\psi}\psi\rangle^2\right]
\label{eqn:chi}
\end{equation}
where the $\langle\rangle$ indicates an average over the ensemble of gauge
configurations and $V$ is the space-time volume of the lattice. Since the
fermion functional integrals have already been performed at this stage, this
quantity is actually the disconnected part of the chiral susceptibility. Since
we use stochastic estimators for $\bar{\psi}\psi$, we obtain an unbiased
estimator for this quantity by using several independent estimates for each
configuration (5, in fact). Our estimate of $(\bar{\psi}\psi)^2$ is then given
by the average of the (10) estimates which are `off diagonal' in the noise.

\section{Simulations of lattice QCD with 3 sextet quarks.} 

\subsection{$\bm{N_t=6}$}

We simulate lattice QCD with 3 flavours of colour-sextet quarks on a 
$12^3 \times 6$ lattice, using quark masses $m=0.02$, $m=0.01$ and $m=0.005$
which should be adequate to enable extrapolation to the chiral limit. Because
we are primarily interested in the chiral transition we concentrate on the
range $5.5 \le \beta \le 7.0$. Our earlier simulations covered this range with
$\beta$ values spaced by $0.1$, using 10,000 length-1 trajectories for each
$(\beta,m)$. Since we have determined that 
$\beta_\chi(N_t=8) - \beta_\chi(N_t=6) \sim 0.1$ this spacing is inadequate to
determine this evolution of the coupling at the chiral transition with any
precision. We have therefore extended our simulations at the lowest quark mass
($0.005$), to cover the interval $6.2 \le \beta \le 6.4$, which is close to this
transition, with $\beta$s spaced by $0.02$. In addition, we have increased our
statistics to 100,000 trajectories at each of these $\beta$s. Note that we are
only dealing with the phase with a real positive Wilson Line (Polyakov Loop).

\begin{figure}[htb]
\epsfxsize=6.0in
\epsffile{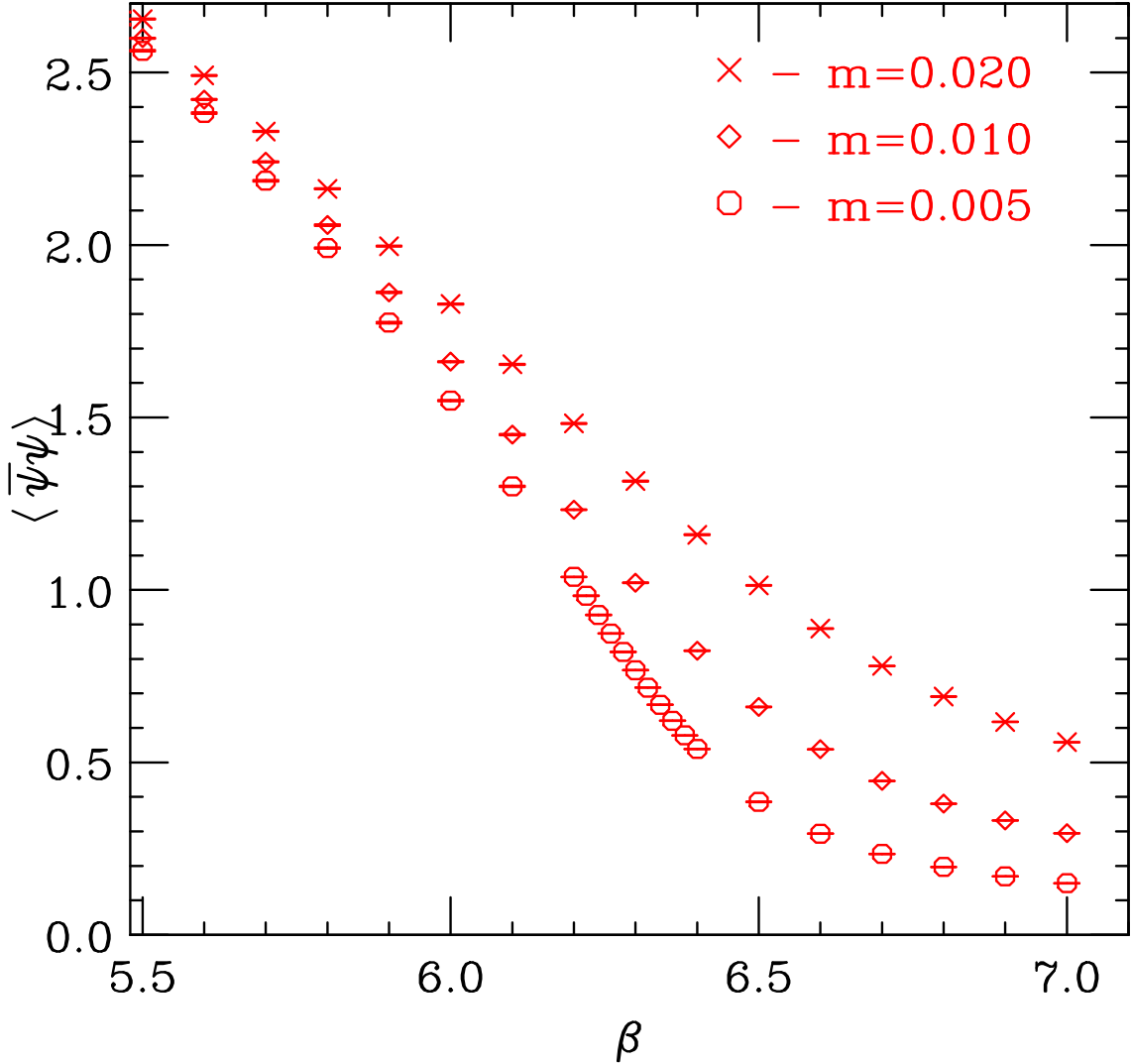}
\caption{$\langle\bar{\psi}\psi\rangle$ as functions of $\beta=6/g^2$ on a
$12^3 \times 6$ lattice, in the state with a real positive Wilson Line, for
$m=0.02,0.01,0.005$.}
\label{fig:rpbp6}
\end{figure}

Figure~\ref{fig:rpbp6} shows the chiral condensates 
($\langle\bar{\psi}\psi\rangle$) as functions of $\beta$ from these simulations.
$\langle\bar{\psi}\psi\rangle$ is normalized to one staggered fermion 
(4 continuum flavours). Note also that these are lattice regularized quantities.
No attempt has been made to subtract the quadratic divergence at $m \ne 0$.
Although these imply that $\langle\bar{\psi}\psi\rangle$ will vanish in the
chiral limit for sufficiently large $\beta$, the dependence on $\beta$ is too
smooth to obtain $\beta_\chi$, the position of the phase transition at which
this vanishing occurs.

\begin{figure}[htb]
\epsfxsize=6.0in
\epsffile{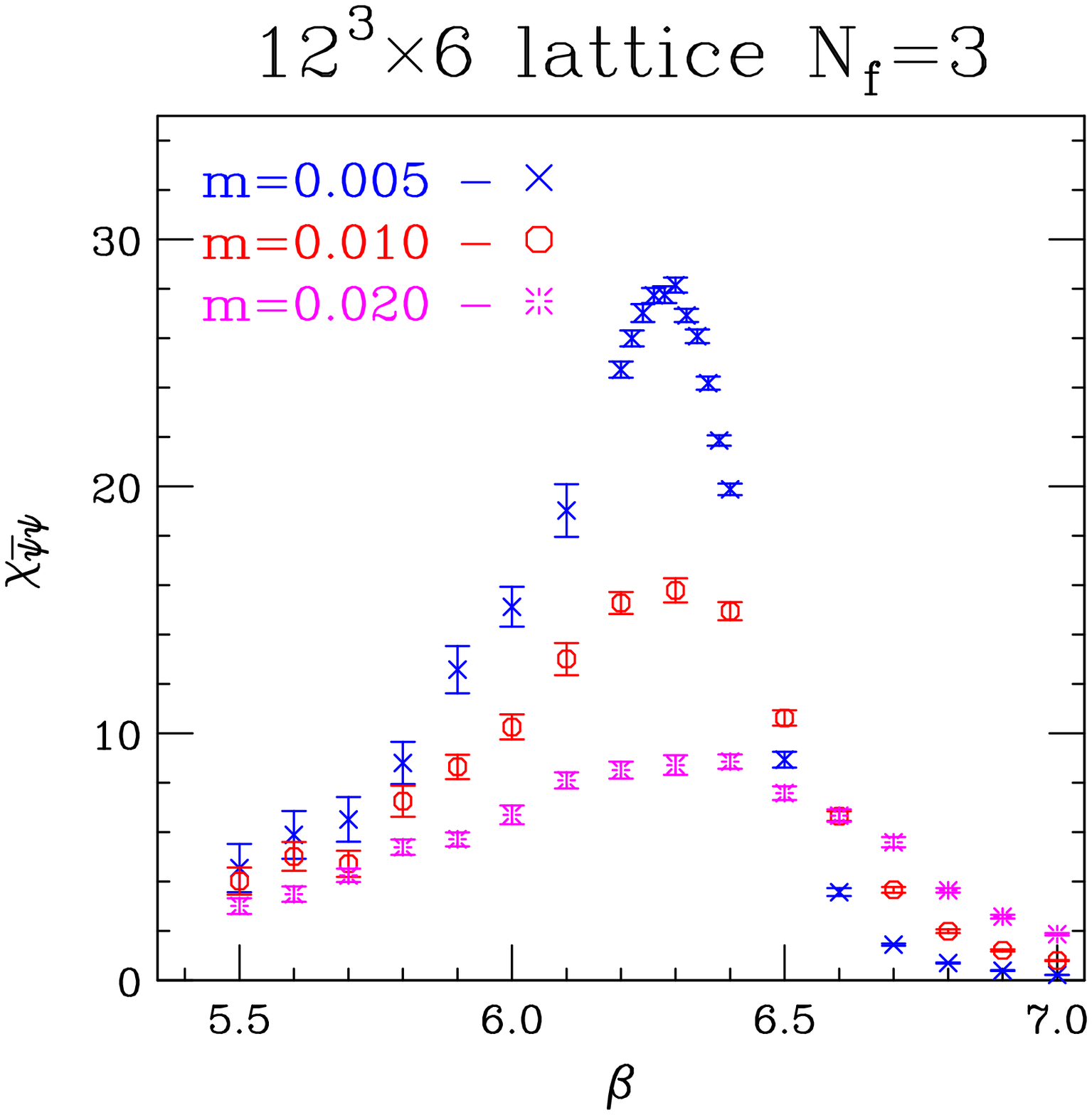}
\caption{Chiral susceptibilities  as functions of $\beta=6/g^2$ on a
$12^3 \times 6$ lattice, in the state with a real positive Wilson Line, for
$m=0.02,0.01,0.005$.}
\label{fig:chipbp6}
\end{figure}

For this reason, we measure the disconnected chiral susceptibility, defined in 
equation~\ref{eqn:chi}, to determine $\beta_\chi$. In the chiral limit, this
quantity diverges at $\beta=\beta_\chi$. For finite $m$ this susceptibility
has a peak, which becomes more pronounced and approaches $\beta_\chi$, as
$m \rightarrow 0$. In practice we have found that the position of the peak
$\beta_{max}$ has little $m$ dependence, and so $\beta_{max}$ for the smallest
$m$ should be a good estimate of $\beta_\chi$. We use 5 stochastic estimators
of $\langle\bar{\psi}\psi\rangle$ at the end of each trajectory, to obtain an 
unbiased estimator for $\chi_{\bar{\psi}\psi}$.

\begin{figure}[htb]
\epsfxsize=6.0in
\epsffile{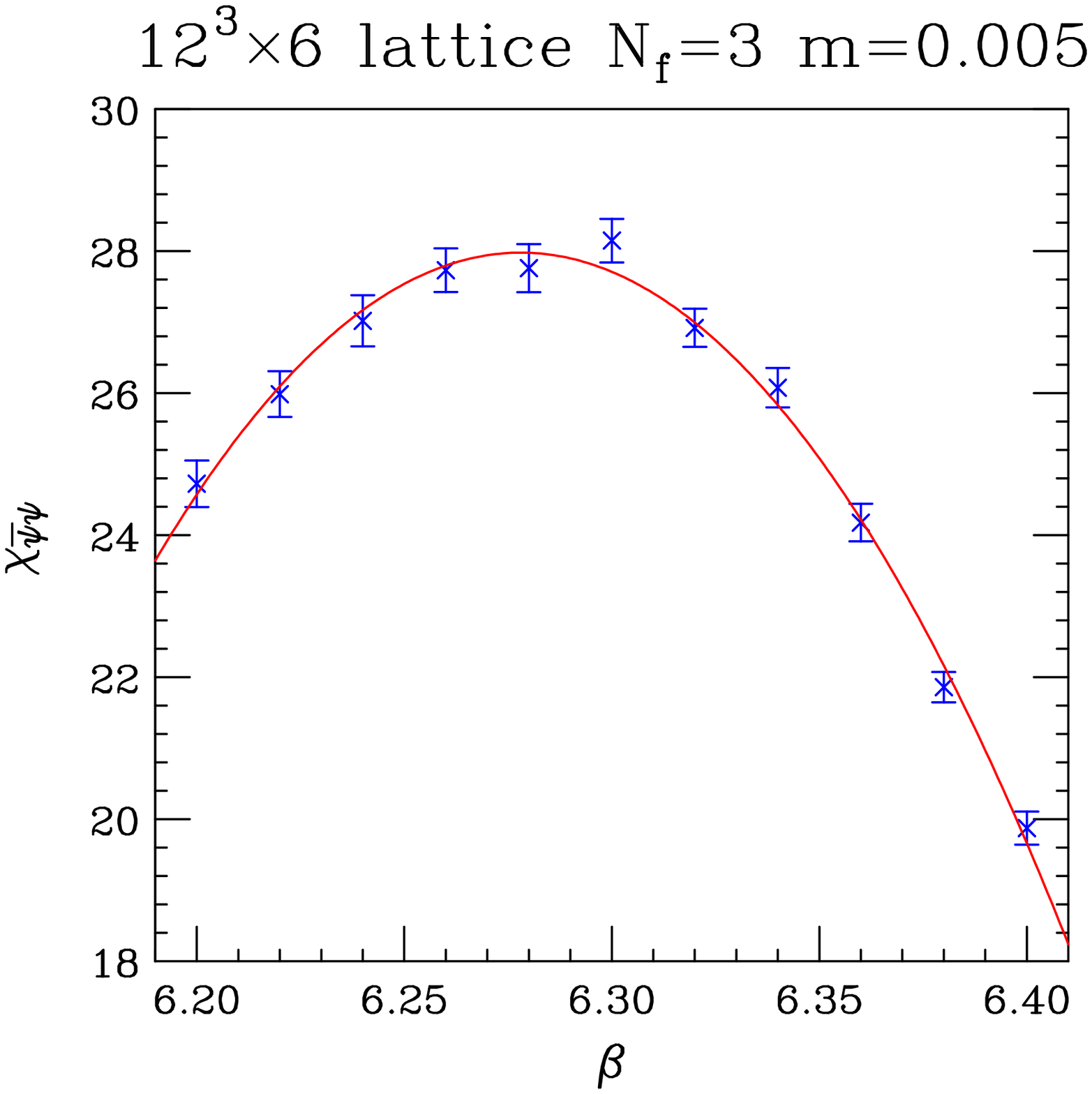}
\caption{Chiral susceptibility for $m=0.005$ as a function of $\beta$ on a
$12^3 \times 6$ lattice, showing the fit to a parabola mentioned in the text.}
\label{fig:chipbp6m005}
\end{figure}

Figure~\ref{fig:chipbp6} shows the susceptibilities measured during these
simulations. The peak in the $m=0.005$ susceptibility is very pronounced, that
for the two higher masses somewhat less so. However, it is clear that the
position of the peak ($\beta_{max}$) has little $m$ dependence. We tried using
Ferrenberg-Swendsen reweighting to obtain precise estimates for the value of
$\beta_{max}$. However, despite the fact that the $\beta$ values in the 
transition region are sufficiently close that there is significant overlap
between plaquette distributions for neighbouring $\beta$s, we had little 
success. Therefore, to make maximal use of our data, we fit the susceptibilities
for $\beta$ values close to the peak to a parabola, namely:
\begin{equation}
\chi_{\bar{\psi}\psi} = a - b (\beta-\beta_{max})^2
\end{equation}
For $m=0.005$, a fit over the range $6.2 \le \beta \le 6.4$ yielded
$\beta_{max}=6.278(2)$ with $\chi^2/d.o.f=0.85$. This fit is shown in 
figure~\ref{fig:chipbp6m005}. It is stable on removal of the point at 
$\beta=6.2$ or that at $\beta=6.4$ from our fit. We also performed fits to our
$m=0.01$ and $m=0.02$ `data'. Since the measurements used in these fits are
spaced by $0.1$ in $\beta$ and the peaks are broad and less pronounced, we 
expect these fits to be less reliable than those for $m=0.005$. For $m=0.01$, 
a fit over $6.0 \le \beta \le 6.5$ yields $\beta_{max}=6.261(8)$ with
$\chi^2/d.o.f=2.1$ while a fit over $6.1 \le \beta \le 6.5$ yields
$\beta_{max}=6.278(9)$ with $\chi^2/d.o.f=1.2$. Finally for $m=0.02$, a fit
over $5.9 \le \beta \le 6.6$ gives $\beta_{max}=6.286(8)$ with
$\chi^2/d.o.f=0.63$, while one over $6.0 \le \beta \le 6.6$ gives
$\beta_{max}=6.294(10)$ with $\chi^2/d.o.f=0.57$. Thus we see that the
position of the peak is almost independent of $m$ which justifies taking the
$m=0.005$ value of $\beta_{max}$ as $\beta_\chi$.

\subsection{$\bm{N_t=8}$}

We perform simulations on $16^3 \times 8$ lattices at $m=0.01$ and $m=0.005$. 
For $m=0.01$ we simulate for $\beta$s in the range $5.4 \le \beta \le 7.0$ at
intervals of $0.1$ in $\beta$, using runs of 10,000 trajectories at each 
$\beta$. For $m=0.005$ we perform simulations for $\beta$s in the range
$5.8 \le \beta \le 7.0$, again at $\beta$ separations of $0.1$ with 10,000
trajectories at each $\beta$, except in the range $6.28 \le \beta \le 6.5$,
where we simulate at $\beta$s spaced by $0.02$ with 100,000 trajectories for
each $\beta$.

\begin{figure}[htb]
\epsfxsize=4.0in
\epsffile{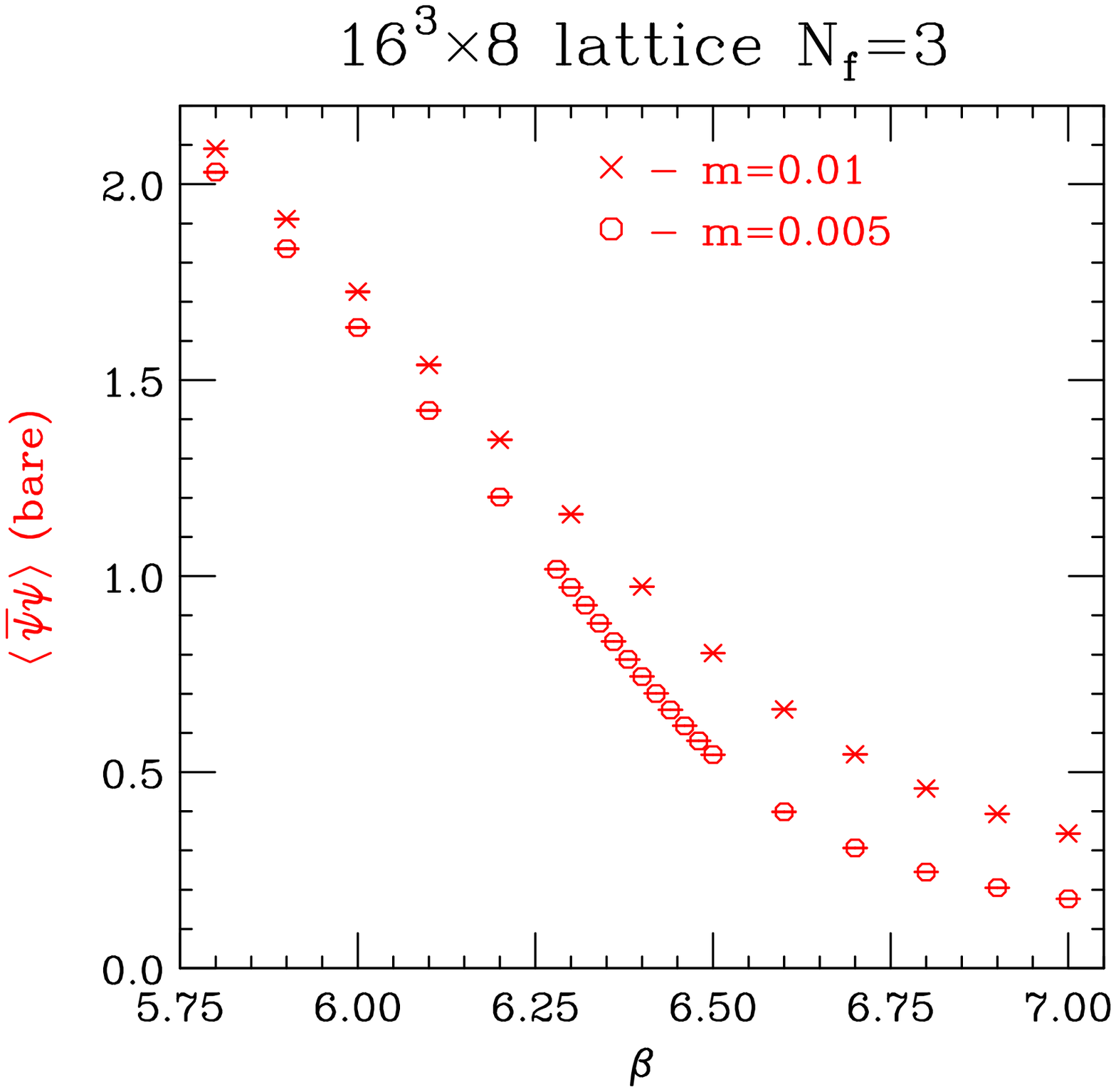}
(a)
\epsfxsize=4.0in
\epsffile{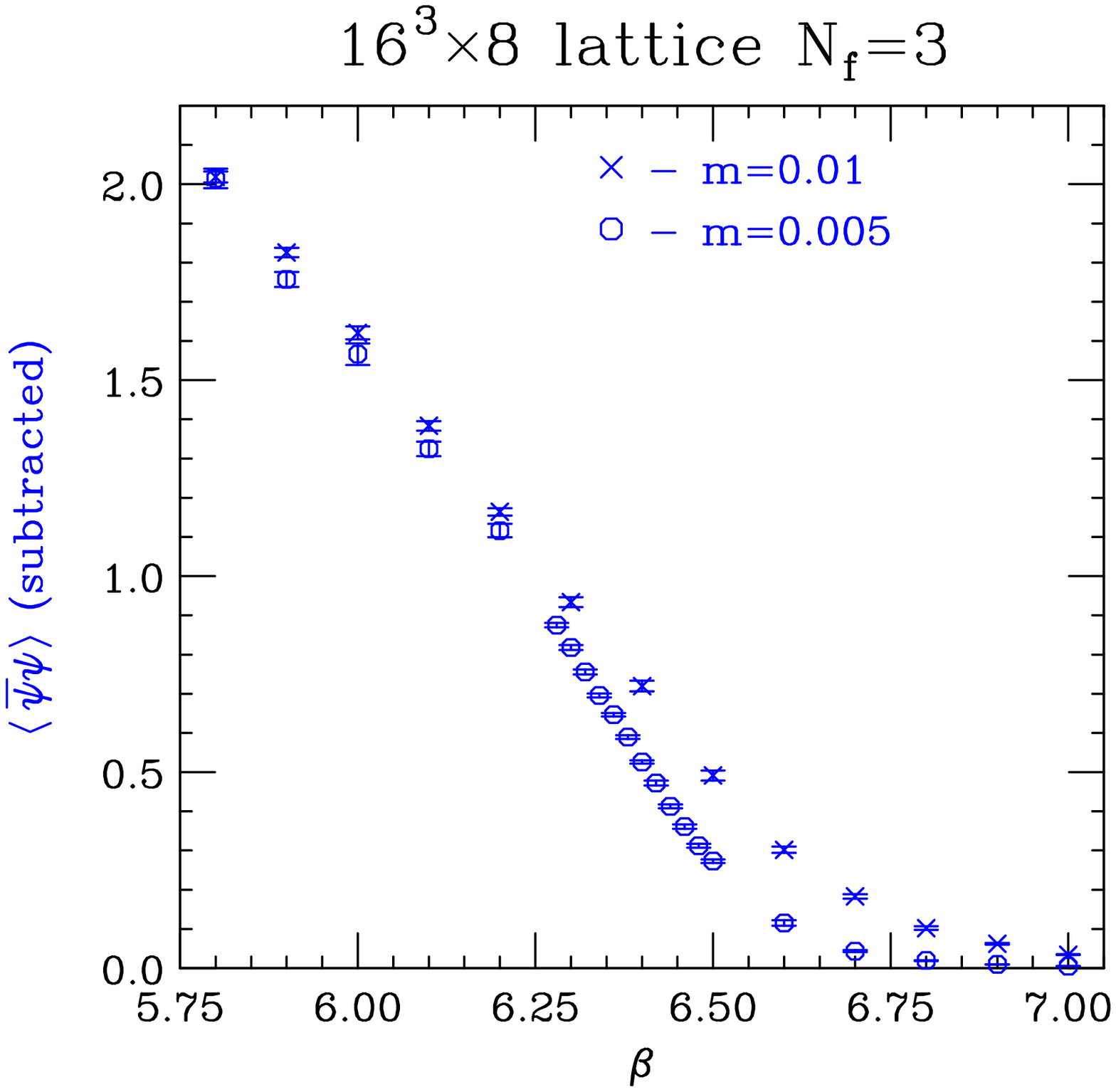}
(b)
\caption{a) $\langle\bar{\psi}\psi\rangle$ as functions of $\beta=6/g^2$ on a
$16^3 \times 8$ lattice, in the state with a real positive Wilson Line, for
$m=0.01,0.005$. \newline
b) $\langle\bar{\psi}\psi\rangle$ on a $16^3 \times 8$ lattice,
subtracted using the valence subtraction used by the Lattice Higgs
Collaboration.}
\label{fig:rpbp8}
\end{figure}

Figure~\ref{fig:rpbp8} shows the chiral condensates measured in these 
simulations. Part~a is the unsubtracted condensate, while part~b is the same
condensates after subtracting much of the quadratic divergence of the 
unsubtracted condensate using the method favoured by the Lattice Higgs 
Collaboration.
\begin{equation}
\langle{\bar{\psi}\psi}\rangle_{sub} = \langle{\bar{\psi}\psi}\rangle 
-\left(m_V\frac{\partial}{\partial m_V}\langle{\bar{\psi}\psi}\rangle\right)
                                                                  _{m_V=m}
\end{equation}
where $m_V$ is the valence quark mass. Although the subtracted condensate
indicates that it will vanish in the chiral limit for large enough $\beta$, more
clearly than the unsubtracted condensate, it still does not yield a precise 
estimate of $\beta_\chi$. Hence we again turn to the chiral susceptibility to
obtain an accurate estimate of $\beta_\chi$.

\begin{figure}[htb]
\epsfxsize=6.0in
\epsffile{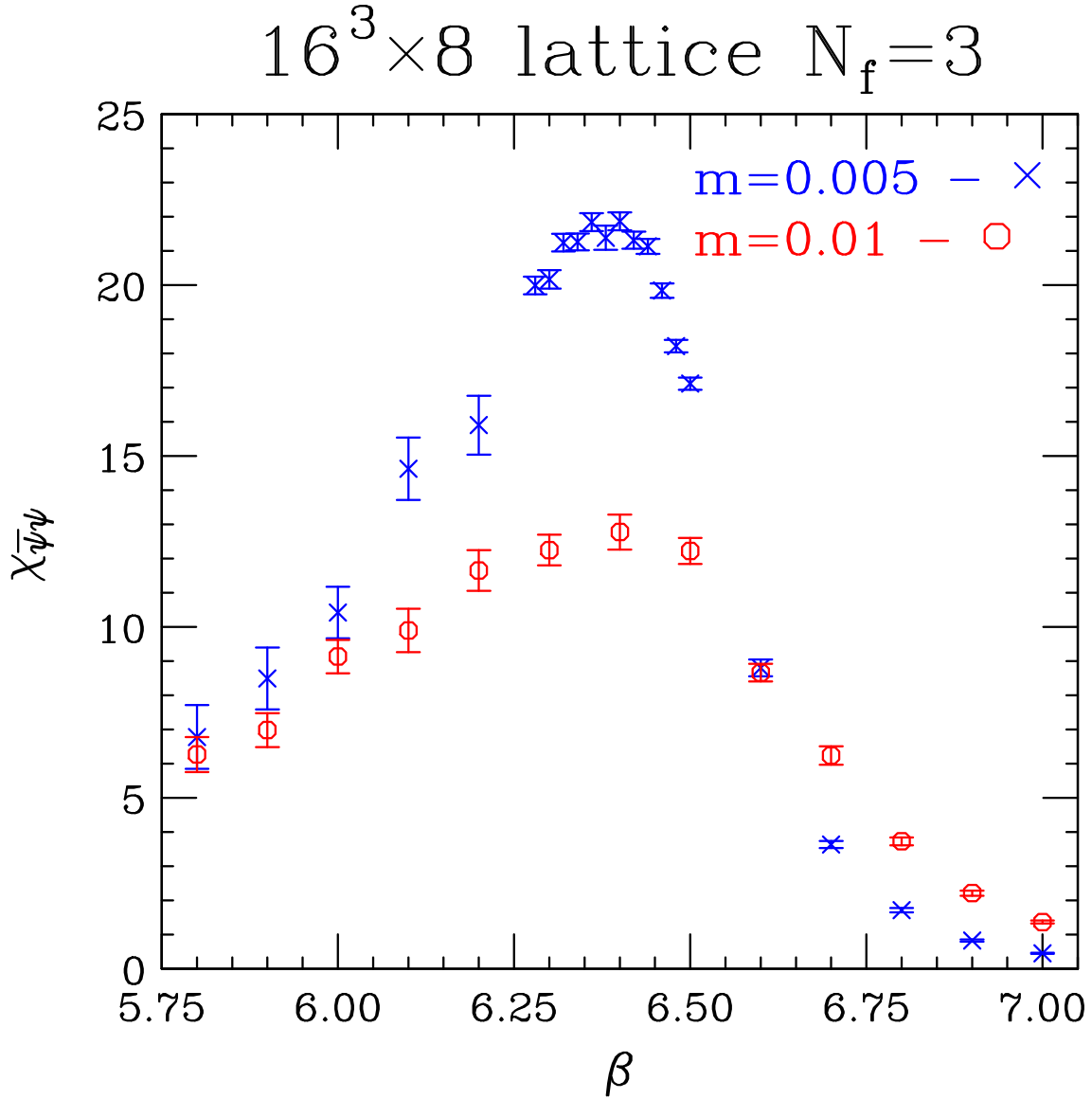}
\caption{Chiral susceptibilities  as functions of $\beta=6/g^2$ on a
$16^3 \times 8$ lattice, in the state with a real positive Wilson Line, for
$m=0.01,0.005$.}
\label{fig:chipbp8}
\end{figure}

\begin{figure}[htb]
\epsfxsize=6.0in
\epsffile{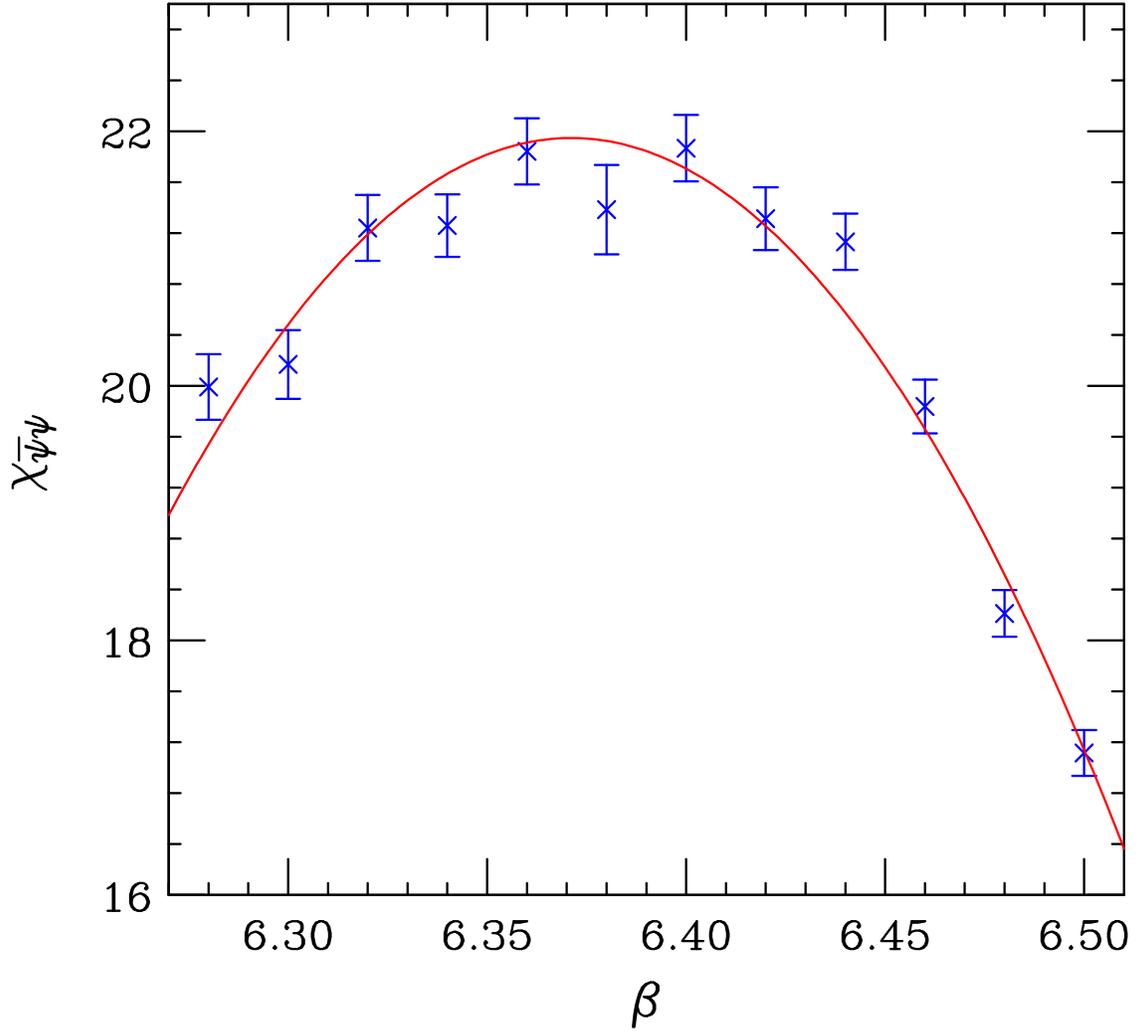}
\caption{Chiral susceptibility as a function of $\beta=6/g^2$ on a
$16^3 \times 8$ lattice, in the state with a real positive Wilson Line, for
$0.005$, showing the parabolic fit mentioned in the text.}
\label{fig:chipbp8m005}
\end{figure}

Figure~\ref{fig:chipbp8} shows the (disconnected) chiral susceptibilities from
our $16^3 \times 8$ simulations at $m=0.005$ and $m=0.01$. Both masses show
peaks, and as expected, the peak at the lower mass is more pronounced. To make
best use of the `data' close to the transition, we again fit this to a
parabola. Fitting over the range $6.28 \le \beta \le 6.5$ where each of the
points represents 100,000 trajectories (actually 80,000, since we discard the
first 20,000 for equilibration), we find the position of the peak to be at
$\beta_{max}=6.371(3)$, while the fit has $\chi^2/d.o.f.=2.2$. This fit is
plotted over the `data' in figure~\ref{fig:chipbp8m005}. Not surprisingly,
this fit is less stable than the fit to the $N_t=6$ susceptibilities. Removing
the point at $\beta=6.28$ changes the peak to $\beta_{max}=6.376(3)$ while
improving the quality of the fit to $\chi^2/d.o.f.=1.6$. We therefore give
$\beta_{max}=6.37(1)$ as our best fit. A fit to the $m=0.01$ `data' over the
range $6.1 \le \beta \le 6.6$ yields $\beta_{max}=6.34(2)$ with 
$\chi^2/d.o.f.=2.5$. Removing the point at $\beta=6.1$ moves the peak to 
$\beta_{max}=6.36(2)$ with almost no change in $\chi^2/d.o.f.$. Again, we 
consider $\beta_{max}$ at $m=0.005$ as our best estimate of $\beta_\chi$. 

The increase in $\beta_\chi$ from $N_t=6$ to $N_t=8$ is thus:
\begin{equation}
\beta_\chi(N_t=8) - \beta_\chi(N_t=6) = 6.37(1) - 6.278(2) = 0.09(1)
\end{equation}
This is to be compared with the increase from $N_t=4$ to $N_t=6$
\begin{equation}                                             
\beta_\chi(N_t=6) - \beta_\chi(N_t=4) \sim 0.3
\end{equation}
If we were still in the same scaling regime, one might expect that
\begin{equation}
{\beta_\chi(N_t=8) - \beta_\chi(N_t=6) \over 
\beta_\chi(N_t=6) - \beta_\chi(N_t=4)} \sim {\ln(8/6) \over \ln(6/4)} 
\approx 0.71
\end{equation}
The fact that the increase from $N_t=6$ to $N_t=8$ is somewhat smaller indicates
that we are emerging from the strong coupling domain. However, there is still
no sign that $\beta_\chi$ is approaching the expected non-zero constant value,
expected for a bulk transition. If, however, this 3-flavour theory is QCD-like,
the 2-loop $\beta$-function would predict (equation~\ref{eqn:deltabeta})
\begin{equation}                                                              
\beta_\chi(N_t=8) - \beta_\chi(N_t=6) = 0.002 \!-\!\!\!-\, 0.003
\end{equation}
which is much smaller than what is observed. It should be noted that
the reason this is so small is that $\beta_\chi(N_t=8)$ is still quite close to
the second zero of the $\beta$-function. For this reason 2-loop perturbation 
theory should not be trusted. However, the 2-loop $\beta$-function never
becomes very large, so that the maximum possible change in $\beta_\chi$
between $N_t=6$ and $N_t=8$, which it predicts, is only $\approx 0.022$. This
bound is true to all orders, provided we are on the weak coupling side of all
non-trivial minima of the $\beta$-function. Hence, although we have evidence
of emergence from the strong-coupling domain, we are not yet able to access
the large $N_t$ limit.

\section{Discussion and conclusions}

We perform simulations of lattice QCD with 3 light colour-sextet quarks, to
compare and contrast its behaviour with that of the 2-flavour version, a
candidate walking-Technicolor theory. Finite-temperature simulations are
performed at $N_t=6$ and $N_t=8$ to determine $\beta_\chi$ the coupling
$\beta=6/g^2$ at the chiral-symmetry restoration phase transitions. In
particular we are looking for evidence that $\beta_\chi$ tends to a finite
limit as $N_t\rightarrow\infty$, since it is expected that this theory is a
conformal field theory, where the chiral transition is a bulk transition.

Our simulations show that between $N_t=6$ and $N_t=8$, $\beta_\chi$ increases
by $0.09(1)$. Although this indicates that by $N_t=8$, $\beta_\chi$ is no longer
in the strong-coupling domain, it does not yet indicate that the chiral 
transition is a bulk phase transition. $\beta_\chi$ also fails to evolve as
would be predicted by asymptotic freedom for a finite temperature transition.
Hence we will need to perform simulations at larger $N_t$. (Simulations at 
$N_t=12$ have been started.)

Table~\ref{tab:trans} gives the positions of the deconfinement transitions
($\beta_d$) and the chiral transitions ($\beta_\chi$) for $N_t=4,6,8$ from this
and previous studies. (Here we note that, since we have only used a single
spatial volume for each $N_t$, lack of significant finite volume corrections is
only an assumption, based on previous experience, which suggests that $N_s=2N_t$
is adequate.)
\begin{table}[h]
\centerline{
\begin{tabular}{|c|l|l|}
\hline
$N_t$  & \multicolumn{1}{c|}{$\beta_d$} & \multicolumn{1}{c|}{$\beta_\chi$} \\
\hline
4              &$\;$5.275(10)$\;$  &$\;$6.0(1)  $\;$            \\
6              &$\;$5.375(10)$\;$  &$\;$6.278(2)$\;$            \\
8              &$\;$5.45(10) $\;$  &$\;$6.37(1) $\;$            \\
\hline
\end{tabular}
}
\caption{$N_f=3$ deconfinement and chiral transitions for $N_t=4,6,8$. In each
case we have attempted an extrapolation to the chiral limit.}
\label{tab:trans}
\end{table}

In the simulations described in this paper, we have worked in the phase where
the Wilson Line (Polyakov Loop) is real and positive. These Wilson Lines are
large close to the chiral transitions, and show little if any indications of
this transition.

\section*{Acknowledgements}

DKS is supported in part by the U.S. Department of Energy, Division of High
Energy Physics, Contract DE-AC02-06CH11357. 

This research used resources of the National Energy Research Scientific
Computing Center, which is supported by the Office of Science of the U.S.
Department of Energy under Contract No. DE-AC02-05CH11231. In particular it
used the IBM Dataplex, Carver and the Cray XE6, Hopper. In addition this 
research used the Cray XT5, Kraken at NICS under XSEDE Project Number: 
TG-MCA99S015. Finally use was made of the Fusion and Blues clusters belonging
to Argonne's LCRC.

DKS would like to thank the CSSM in the University of Adelaide Physics 
Department for hospitality during part of this research.


\begin{thebibliography}{99}


\bibitem{Weinberg:1979bn}
  S.~Weinberg,
  Phys.\ Rev.\  D {\bf 19}, 1277 (1979).

\bibitem{Susskind:1978ms}
  L.~Susskind,
  Phys.\ Rev.\  D {\bf 20}, 2619 (1979).


\bibitem{Holdom:1981rm}
  B.~Holdom,
  Phys.\ Rev.\  D {\bf 24}, 1441 (1981).

\bibitem{Yamawaki:1985zg}
  K.~Yamawaki, M.~Bando and K.~i.~Matumoto,
  Phys.\ Rev.\ Lett.\  {\bf 56}, 1335 (1986).

\bibitem{Akiba:1985rr}
  T.~Akiba and T.~Yanagida,
  Phys.\ Lett.\  B {\bf 169}, 432 (1986).


\bibitem{Appelquist:1986an}
  T.~W.~Appelquist, D.~Karabali and L.~C.~R.~Wijewardhana,
  Phys.\ Rev.\ Lett.\  {\bf 57}, 957 (1986).


\bibitem{Dietrich:2006cm}
  D.~D.~Dietrich and F.~Sannino,
  Phys.\ Rev.\  D {\bf 75}, 085018 (2007)
  [arXiv:hep-ph/0611341].


\bibitem{Kogut:2010cz}
  J.~B.~Kogut, D.~K.~Sinclair,
  Phys.\ Rev.\  {\bf D81}, 114507 (2010).
  [arXiv:1002.2988 [hep-lat]].

\bibitem{Kogut:2011ty} 
  J.~B.~Kogut and D.~K.~Sinclair,
  Phys.\ Rev.\ D {\bf 84}, 074504 (2011)
  [arXiv:1105.3749 [hep-lat]].

\bibitem{Sinclair:2013era} 
  D.~K.~Sinclair and J.~B.~Kogut,
  arXiv:1311.5208 [hep-lat].


\bibitem{Shamir:2008pb}
  Y.~Shamir, B.~Svetitsky and T.~DeGrand,
  Phys.\ Rev.\  D {\bf 78}, 031502 (2008)
  [arXiv:0803.1707 [hep-lat]].

\bibitem{DeGrand:2008kx}
  T.~DeGrand, Y.~Shamir and B.~Svetitsky,
  Phys.\ Rev.\  D {\bf 79}, 034501 (2009)
  [arXiv:0812.1427 [hep-lat]].

\bibitem{DeGrand:2009hu}
  T.~DeGrand,
  Phys.\ Rev.\  D {\bf 80}, 114507 (2009)
  [arXiv:0910.3072 [hep-lat]].

\bibitem{DeGrand:2010na}
  T.~DeGrand, Y.~Shamir and B.~Svetitsky,
  Phys.\ Rev.\  D {\bf 82}, 054503 (2010)
  [arXiv:1006.0707 [hep-lat]].

\bibitem{DeGrand:2012yq} 
  T.~DeGrand, Y.~Shamir and B.~Svetitsky,
  Phys.\ Rev.\ D {\bf 87}, 074507 (2013)
  [arXiv:1201.0935 [hep-lat]].

\bibitem{DeGrand:2013uha} 
  T.~DeGrand, Y.~Shamir and B.~Svetitsky,
  Phys.\ Rev.\ D {\bf 88}, no. 5, 054505 (2013)
  [arXiv:1307.2425].

\bibitem{Fodor:2008hm} 
  Z.~Fodor, K.~Holland, J.~Kuti, D.~Nogradi and C.~Schroeder,
  PoS LATTICE {\bf 2008}, 058 (2008)
  [arXiv:0809.4888 [hep-lat]].

\bibitem{Fodor:2011tw} 
  Z.~Fodor, K.~Holland, J.~Kuti, D.~Nogradi and C.~Schroeder,
  PoS Lattice {\bf 2010}, 060 (2010)
  [arXiv:1103.5998 [hep-lat]].

\bibitem{Fodor:2012uu} 
  Z.~Fodor, K.~Holland, J.~Kuti, D.~Nogradi, C.~Schroeder and C.~H.~Wong,
  PoS Lattice {\bf 2011}, 073 (2011)
  [arXiv:1205.1878 [hep-lat]].

\bibitem{Fodor:2012uw} 
  Z.~Fodor, K.~Holland, J.~Kuti, D.~Nogradi, C.~Schroeder and C.~H.~Wong,
  PoS LATTICE {\bf 2012}, 025 (2012)
  [arXiv:1211.3548 [hep-lat]].

\bibitem{Fodor:2012ty} 
  Z.~Fodor, K.~Holland, J.~Kuti, D.~Nogradi, C.~Schroeder and C.~H.~Wong,
  Phys.\ Lett.\ B {\bf 718}, 657 (2012)
  [arXiv:1209.0391 [hep-lat]].

\bibitem{Fodor:2014pqa} 
  Z.~Fodor, K.~Holland, J.~Kuti, D.~Nogradi and C.~H.~Wong,
  arXiv:1401.2176 [hep-lat].


\bibitem{Appelquist:1988yc}
  T.~Appelquist, K.~D.~Lane and U.~Mahanta,
  Phys.\ Rev.\ Lett.\  {\bf 61}, 1553 (1988).

\bibitem{Sannino:2004qp}
  F.~Sannino and K.~Tuominen,
  Phys.\ Rev.\  D {\bf 71}, 051901 (2005)
  [arXiv:hep-ph/0405209].

\bibitem{Poppitz:2009uq}
  E.~Poppitz and M.~Unsal,
  JHEP {\bf 0909}, 050 (2009)
  [arXiv:0906.5156 [hep-th]].

\bibitem{Armoni:2009jn} 
  A.~Armoni,
  Nucl.\ Phys.\ B {\bf 826}, 328 (2010)
  [arXiv:0907.4091 [hep-ph]].


\bibitem{Ryttov:2007cx}
  T.~A.~Ryttov and F.~Sannino,
  Phys.\ Rev.\  D {\bf 78}, 065001 (2008)
  [arXiv:0711.3745 [hep-th]].

\bibitem{Antipin:2009wr} 
  O.~Antipin and K.~Tuominen,
  Phys.\ Rev.\ D {\bf 81}, 076011 (2010)
  [arXiv:0909.4879 [hep-ph]].


\bibitem{Clark:2006wp}
  M.~A.~Clark and A.~D.~Kennedy,
  Phys.\ Rev.\  D {\bf 75}, 011502 (2007)
  [arXiv:hep-lat/0610047].


\bibitem{Kogut:2011bd} 
  J.~B.~Kogut and D.~K.~Sinclair,
  Phys.\ Rev.\ D {\bf 85}, 054505 (2012)
  [arXiv:1111.3353 [hep-lat]].


\bibitem{Sharpe:2006re}
  S.~R.~Sharpe,
  PoS {\bf LAT2006}, 022 (2006)
  [arXiv:hep-lat/0610094].


\end{thebibliography}
\end{document}